\documentclass[preprint2]{aastex}

\usepackage{graphicx}
\usepackage{natbib}
\usepackage{color}
\usepackage{epstopdf}

\def\deg {{$^{\circ}$}}
\def\wn {{cm$^{-1}$}~}
\def\wnp {{cm$^{-1}$}}
\def\um {{$\mu$m}}
\def\nutwo {{$\nu$$_{2}$}}
\def\nuthree {{$\nu$$_{3}$}}
\def\nufour {{$\nu$$_{4}$}}

\def\ph {{PH$_{3}$}}
\def\ammonia {{NH$_{3}$}}
\def\water {{H$_{2}$O}}
\def\hs {{H$_{2}$S}}
\def\dm {{CH$_{3}$D}}
\def\methane {{CH$_{4}$}}
\def\amhs {{NH$_{4}$SH}}
\def\h {{H$_{2}$}}
\def\arcsec {{$^{\prime \prime}$}}

\shorttitle{Jupiter's Deep Cloud Structure}
\shortauthors{Bjoraker et al.}

\begin{document}

\title{Jupiter's Deep Cloud Structure Revealed Using Keck Observations of Spectrally Resolved Line Shapes}


\author{G. L. Bjoraker}
\affil{NASA/GSFC Code 693, Greenbelt, MD 20771, USA}
\email{gordon.l.bjoraker@nasa.gov}

\author{M. H. Wong}
\affil{Department of Astronomy, University of California, Berkeley, CA 94720-3411, USA}

\author{I. de Pater}
\affil{Department of Astronomy, University of California, Berkeley, CA 94720-3411, USA}

\and

\author{M. \'{A}d\'{a}mkovics}
\affil{Department of Astronomy, University of California, Berkeley, CA 94720-3411, USA}




\begin{abstract}

\emph{Technique:} We present a method to determine the pressure at which significant cloud opacity is present between 2 and 6~bars on Jupiter. We use: a) the strength of a Fraunhofer absorption line in a zone to determine the ratio of reflected sunlight to thermal emission, and b) pressure-broadened line profiles of deuterated methane (\dm)~ at 4.66~\um~ to determine the location of clouds. We use radiative transfer models to constrain the altitude region of both the solar and thermal components of Jupiter's 5-\um~spectrum. \emph{Results:} For nearly all latitudes on Jupiter the thermal component is large enough to constrain the deep cloud structure even when upper clouds are present. We find that Hot Spots, belts, and high latitudes have broader line profiles than do zones. Radiative transfer models show that Hot Spots in the North and South Equatorial Belts (NEB, SEB) typically do not have opaque clouds at pressures greater than 2 bars. The South Tropical Zone (STZ) at 32\deg S has an opaque cloud top between 4 and 5 bars. From thermochemical models this must be a water cloud. We measured the variation of the equivalent width of \dm~ with latitude for comparison with Jupiter's belt-zone structure. We also constrained the vertical profile of \water~ in an SEB Hot Spot and in the STZ. The Hot Spot is very dry for P$<$4.5~bars and then follows the \water~ profile observed by the Galileo Probe. The STZ has a saturated \water~ profile above its cloud top between 4 and 5~bars. \end{abstract}


\keywords{planets and satellites: individual (Jupiter) --- planets and satellites: atmospheres}



\section{INTRODUCTION}

The abundance of \water~in Jupiter's atmosphere is of fundamental importance in understanding the origin of Jupiter, the composition of its clouds, and Jovian dynamics beneath the upper cloud layers. Water was first detected on Jupiter by \citet{larson75} using the Kuiper Airborne Observatory (KAO). \citet{bjoraker86b, bjoraker86a} analyzed both KAO and Voyager IRIS spectra in Jupiter's 5-\um~spectral window to the deep atmosphere, finding a deep water abundance highly depleted with respect to the solar abundance.  The spectrum between 4.5 and 5.4 \um~provides a wealth of information about the gas composition and cloud structure of the troposphere of this giant planet. Jupiter's 5-\um~spectrum is a mixture of scattered sunlight and thermal emission that changes significantly between belts and zones. Jupiter exhibits remarkable spatial structure at 5~\um. Chemical models of Jupiter's cloud structure predict three distinct layers: an \ammonia~ice cloud near 0.5~bars, an \amhs~cloud formed from a reaction of \ammonia~and \hs~at 2 bars, and a massive water ice/liquid solution cloud near 5 or 6 bars, depending on assumptions of composition and thermal structure (see \citet{weidenschilling73} and \citet{wong15}). Thermal emission from the deep atmosphere is attenuated by the variable opacity of one or more of these three cloud layers. Hot Spots, located primarily in the North and South Equatorial Belts (NEB, SEB), exhibit 5-\um~radiances up to 70 times larger than surrounding regions due to a minimum of cloud opacity. They also appear brighter than their surroundings at microwave wavelengths due to a low ammonia abundance \citep{sault04}.

The interpretation of 5-\um~spectra of Jupiter, and the \water~abundance in particular, has been hampered by uncertainties in the pressure level of the lower boundary of the emitting region. Two different models have been proposed.  \citet{bjoraker86b, bjoraker86a}  suggested that thermal radiation at 5~\um~originates from levels as deep as 8 bars, 310 K where unit optical depth in \h~occurs. In contrast, \citet{carlson92} proposed a model in which a massive water-ice cloud establishes the lower boundary near 5 bars, 273 K. Bjoraker et al. fitted KAO and Voyager IRIS spectra of \water~in Jupiter's Hot Spots with a small abundance (4-30~ppm) distributed along a long path (60~km) between 2 and 8~bars. They also measured \water~abundances in the low-flux zone regions using Voyager spectra. The NEB hot spots were found to be depleted in \water~between 2 and 4~bars, but belts, zones, and hot spots could all be fitted by the same \water~profile (4-30~ppm) between 4 and 8~bars. In marked contrast, \citeauthor{carlson92} fitted Voyager Hot-Spot spectra using a much larger mixing ratio (up to 3000~ppm, equivalent to 3$\times$ the solar O/H measured by \citet{asplund09}) confined to a narrow layer (10~km) between 4 and 5~bars immediately above an opaque water cloud. Carlson et al. also examined Voyager spectra of the Equatorial Zone (EQZ) and other regions away from hot spots. All regions required a saturated \water~ profile that increased from 300~ppm at 4 bars to 3000~ppm at 5 bars, although NEB hot spots were sub-saturated in the 2 to 4~bar region.

The Galileo Probe measured water vapor in situ at a single location on Jupiter, but unanswered questions remain about its global abundance. The probe entered Jupiter's atmosphere at 6.5\deg N planetocentric latitude near the southern portion of an NEB hot spot. The Galileo Probe Mass Spectrometer \citep{niemann92} found water increasing with depth, from an upper limit of 0.8~ppm at 2.7~bar, to measured mole fractions of 40$\pm$13~ppm at 11.0-11.7~bar and 420$\pm$140~ppm at 18-21~bar (\citet{niemann98} and \citet{wong04}).  The deepest value corresponds to 0.45$\times$ solar O/H. Ground-based imaging has shown that 5-\um~hot spots, such as the one entered by the Galileo probe, cover less than 1\% of the surface area of Jupiter \citep{orton96}.  Thus, the \water~abundances observed by the probe may be characteristic of all or most hot spots, but they are probably not representative of Jupiter as a whole, especially since a range of other indirect studies (lightning flash depths, the tropospheric CO abundance, and discrete clouds at $P \ge 4$~bar) point to solar or supersolar water abundances \citep{wong08}. Since the Galileo probe found that carbon, sulfur, and nitrogen were enriched by $\sim$4$\times$ solar \citep{wong04}, oxygen may also be enhanced by the same amount. Some formation models require $\sim$10$\times$ solar O/H in order to trap Jupiter's volatiles inside cages of water ice clathrates \citep{hersant04}. The Juno mission, scheduled to begin orbiting Jupiter in July 2016, should answer many of these questions. The Microwave Radiometer will measure water vapor below Jupiter's clouds to determine the O/H ratio \citep{janssen05}. Interpretation of these data may not be straightforward, however, due to e.g., the small microwave absorptivity of \water~ gas compared with \ammonia~ (see \citet{depater05} for details). A complementary way to determine the deep \water~ abundance, therefore, is highly desirable. The Jovian Infrared Auroral Mapper (JIRAM) will acquire near-infrared spectra of Jupiter, including the 5-micron window \citep{adriani14}.

Water ice has been detected in isolated regions where active convection lofted the ice well above its condensation level \citep{simon00}, but water clouds are generally hidden by overlying \ammonia~ and \amhs~ clouds (e.g. \citet{sromovsky10}). Only very thin clouds were found in the Galileo Probe Hot Spot \citep{ragent98}, consistent with condensation via weak turbulent updrafts within the descending branch of an equatorially-trapped Rossby wave (\citet{friedson05}, \citet{wong15}).  Models of NIMS spectra by \citet{nixon01} include water clouds in at least some Hot Spots, and \citet{roos04} showed that NIMS spectra of Hot Spots cannot rule out water clouds whose opacity is entirely restricted to $P>5$~bar. Evidence for water clouds in Jupiter's zones is also ambiguous. \citet{drossart98} compared dayside and nightside NIMS spectra of low-flux regions in the EQZ. A saturated \water~profile above an opaque water cloud at 5~bars provides a satisfactory fit to these data. At the spectral resolution of IRIS (4.3~\wnp) and NIMS (10~\wnp) we can only retrieve a column abundance of \water~above an assumed lower boundary, which can be either a water cloud or opacity due to \h. Thus, we simply cannot tell whether or not Hot Spots or zones have water clouds. This problem will also affect the interpretation of 5-\um~spectra from JIRAM on Juno, which has the same spectral resolution as Voyager/IRIS \citep{adriani14}.

In Section 2 we present ground-based observations of Jupiter's belts and zones that have sufficient spectral resolution to resolve line shapes. We demonstrate that we can derive the cloud structure at $P>2$~bars even when higher-altitude clouds greatly attenuate the thermal flux from the deep atmosphere. We used line shapes  to detect water clouds and to determine the pressure at which these clouds become optically thick. This resolves the ambiguity of whether the lower boundary of the 5-\um~ line formation region is due to \h~ or due to opaque clouds. This, in turn, will yield more accurate gas abundances for use in constraining models of Jupiter's origin and in understanding the dynamics of the atmosphere beneath the upper clouds.

\section{OBSERVATIONS}\label{obs}


Five-micron spectra of Jupiter were acquired using NIRSPEC on the Keck 2 telescope on March 11, 2014.  NIRSPEC is an echelle spectrograph with 3 orders dispersed onto a 1024x1024 InSb array at 5 \um~ at our selected grating/cross-disperser settings of 60.48 / 36.9 \citep{mcLean98}.  A 0.4\arcsec $\times$ 24\arcsec slit was aligned north-south on the central meridian of Jupiter, resulting in spectra with a resolving power of 20,000. NIRSPEC has an advantage over instrumentation on other telescopes (for example CSHELL on the IRTF) due to the fact that 3 echelle orders at 5~\um~ are placed on the detector array. Thus, each pixel along the slit corresponding to different  latitudes on Jupiter has simultaneous spectra at 4.6, 5.0, and 5.3~\um. In this paper we will focus on Order 16 covering 2131-2165 \wn (4.62-4.69 \um) and Order 15 which covers 1999-2031 \wn (4.92-5.00 \um). Jupiter subtended 41\arcsec~ and the geocentric Doppler shift was 26.3 km/sec.  The water vapor column above Mauna Kea was 2 precipitable mm derived from fitting telluric lines in the Jupiter spectra. We did not use stellar spectra for flux calibration or atmospheric transmission because the humidity doubled between the time of the Jupiter and stellar observations. The flux calibration is described in Section 3. Fig. 1 shows an image of Jupiter using the SCAM guide camera on NIRSPEC. Although the spectroscopy was performed at 5-\um, the guide camera works at shorter wavelengths. A K-prime filter centered at 2.12 \um~ was used to obtain sufficient contrast to separate the bands of variable haze reflectivity overlying Jupiter's belts and zones. Two slit positions were required to obtain pole-pole spectra of Jupiter. In Fig.~1 spatial pixels that exhibit maxima in flux at 4.66 \um~ are shown in red, locations that exhibit narrow line profiles at 4.66 \um~ are shown in blue, and regions that have broad pressure-broadened line profiles are shown in green. We focus on two spatial locations with characteristically different spectra: Region A in the South Tropical Zone (STZ) at 32\deg S, and Region B, a Hot Spot in the SEB at 17\deg S.  We developed radiative transfer models to calculate synthetic spectra for regions A and B, as described below.

\begin{figure}[!ht]
\begin{center}
\begin{tabular}{ll}
\hspace{-0.12in}
	 \includegraphics[width=2.95in]{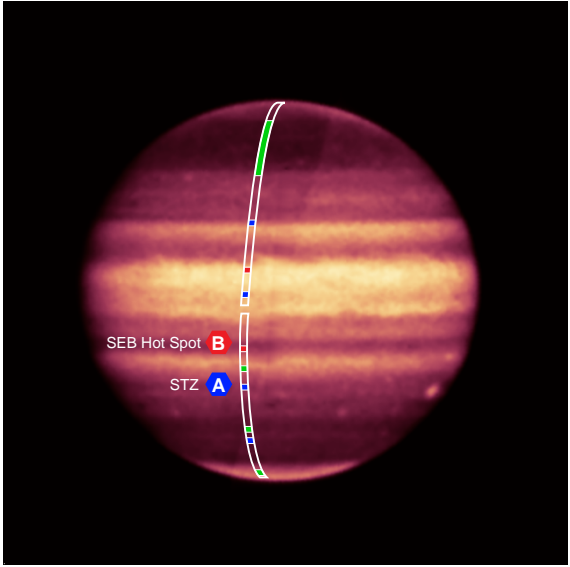}
\end{tabular}	\vspace{-0.1in}
   \caption {\footnotesize {Keck/SCAM image of Jupiter at 2.12~\um~shows 2 NIRSPEC slit positions, one in the northern, and one in the southern hemisphere. Curvature is due to navigating positions onto an image taken a few minutes earlier. Red pixels denote Hot Spots. Blue pixels denote positions of minima in deuterated methane (\dm)~ equivalent width and thus candidates for opaque clouds. Green regions denote maxima in \dm~ equivalent width and thus regions without opaque clouds. Radiative transfer models were used to calculate synthetic spectra for the zone marked A and the SEB Hot Spot labeled B.}}\vspace{-0.2in}
   \label{fig1}
\end{center}  
\end{figure}   

The band pass at 4.66 \um~ contains absorption lines of deuterated methane (\dm), phosphine (\ph), and \water. Methane and its isotopologues do not condense, and they are not destroyed photochemically, in Jupiter's troposphere. We therefore assume that \methane~ and \dm~ have a constant mixing ratio with respect to \h~ in Jupiter's troposphere, which means that variations in the strength and shape of \methane~ and \dm~lines between belts and zones on Jupiter should be due to changes in cloud structure, not gas concentration. \citet{bjoraker02} measured the spatial variation of the weak \nuthree - \nufour~ band of \methane~at 5.18~\um~in an attempt to derive cloud structure. Unfortunately, they were unable to calculate synthetic spectra that fit this feature. This is possibly due to inaccurate or incomplete spectroscopic parameters such as line strengths and broadening parameters. Spectroscopic parameters for the much stronger \nutwo~fundamental band of \dm~at 2200~\wnp, or 4.5~\um, are well known \citep{nikitin97} and as shown below we now are able to derive cloud structure and spatial variations therein.
\begin{figure}[!ht]
\begin{center}
\begin{tabular}{ll}
\hspace{-0.15in}
	 \includegraphics[width=3.0in]{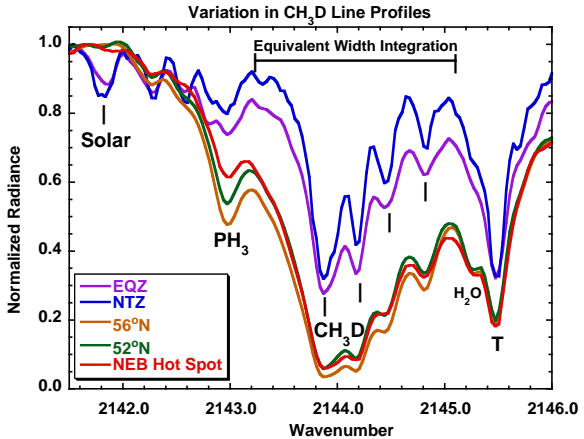}
\end{tabular}	\vspace{-0.1in}
   \caption {\footnotesize {Jupiter spectra at 4.66~\um~ (NIRSPEC order 16) show dramatic variation with latitude in the strength and width of \dm~ lines. Four absorption features are denoted by vertical lines. Narrow \dm~ lines are observed in the Equatorial and North Tropical Zones (EQZ, NTZ), but they are much broader in Hot Spots and at high northern latitudes. A phosphine (\ph)~line and a Doppler-shifted \water~ line are shown. Fraunhofer lines (CO in the Sun) are only observed in low-flux zones. T denotes telluric lines.}}\vspace{-0.2in}
   \label{fig2}
\end{center}  
\end{figure}   

Fig. 2 shows spectra at 4.66 \um~ (2144 \wnp, NIRSPEC order 16) of 5 regions in Jupiter's northern hemisphere that exhibit either a maximum in flux (an NEB Hot Spot at 8.5\deg N), minima in the equivalent width of \dm~ (the EQZ at 0.5\deg N and NTZ at 23\deg N), or local maxima in \dm~ equivalent width. A horizontal bar indicates the limits that we chose for numerical integration of the equivalent width that includes six absorption lines of \dm~(blended into four at this spectral resolution). All spectra are normalized to 1.0 at 2141.6~\wn to facilitate comparison of line shapes, since the radiance of the NEB Hot Spot is 70 times that of the NTZ. A telluric \water~ line, its Doppler-shifted counterpart, a \ph~ feature, and a Fraunhofer line due to CO in the Sun are also shown. Note that the Fraunhofer line is observed only in low-flux zone regions such as the EQZ and NTZ. In Sections 3 and  4 we describe how the strength of this feature can be used to constrain cloud models. The equivalent width of the set of \dm~ lines at 2144~\wn varies dramatically from belt to zone. Hot Spots and belts exhibit broad pressure-broadened line profiles, while zones have much narrower features that are resolved as 4 distinct \dm~ absorption lines. Note that an isolated \dm~ line would be spectrally resolved at a resolving power of 20,000 (0.11~\wn resolution). This is because molecules such as \dm~ typically have broadening coefficients of 0.06~\wnp/atm and the line formation region on Jupiter at 4.66~\um~ takes place at pressures greater than 2 bars.

\section{DATA ANALYSIS AND RESULTS}

Absolute flux calibration was performed for both NIRSPEC orders 15 and 16 by smoothing the spectrum of the NEB Hot Spot at 8.5\deg N to 4.3 \wn resolution, dividing by the transmittance of the Earth's atmosphere above Mauna Kea, and scaling the radiance of the resulting spectrum to an average of the 4 hottest spectra of Jupiter's NEB observed by Voyager IRIS in 1979.

In addition to pressure-broadened line profiles, the spectrum at 4.66~\um~ provides valuable information on the ratio of reflected sunlight to thermal emission on Jupiter, which is critical for understanding Jupiter's cloud structure. We compared the equivalent width of the Fraunhofer line at 2141.8~\wn with its measured value in the Sun using data from the Atmospheric Trace Molecule Spectroscopy (ATMOS) experiment which flew on the space shuttle Challenger in 1985 \citep{farmer89, farmer94}. Solar Fraunhofer lines in spectra of Jupiter arise from regions with a significant fraction of reflected sunlight, such as locations with thick high clouds. Fraunhofer lines are not observed in regions that are dominated by thermal emission, such as Hot Spots. We selected a zone at 32\deg S (marked``A" on Figures 1, 4, and 5) for further study based on the strength of this particular Fraunhofer line.  It is 43\% as strong as in the Sun; thus, 57\% of the flux consists of thermal emission originating in the deep atmosphere that has been attenuated by one or more cloud layers before escaping to space. This thermal flux preserves the broad line profiles of \dm~ caused by collisions with \h~ and He in its line formation region at the deepest levels probed. We excluded the spectrum of the NTZ shown in Fig. 2 for studies of the deep atmosphere because the Fraunhofer line is 95\% as strong as in the Sun; thus, there is insufficient thermal flux to constrain the deep cloud structure at 23\deg N. We also modeled the SEB Hot Spot at 17\deg S and 208\deg W (marked "B" on Figures 1, 4, and 5), which had the highest radiance in our entire dataset. The noise level was measured using the standard deviation of the number of counts in the spectral pixel corresponding to a saturated telluric water line in each order. This was evaluated over 11 spatial pixels over the lowest flux region (the STZ) and also for 11 spatial pixels off of Jupiter. Both gave the same result. In Order 16 the signal to noise ratio (S/N) of the SEB Hot Spot at 2141.6~\wn was 2500; the S/N of the zone at 32\deg S was 48. In Order 15 where strong \water~ lines occur, the S/N of the same Hot Spot was 1900 at 2012.2~\wn; the S/N of the STZ was 35.

Synthetic spectra were calculated using the Spectrum Synthesis Program (SSP) radiative transfer code as described in \citet{kunde74}.  The input temperature profile was obtained from the Galileo Probe (Seiff et al. 1998). Line parameters for \dm~ and other 5-\um~ absorbers are from GEISA 2003 \citep {husson05}. Parameters for \dm-\h~and \dm-He broadening have been measured in the lab \citep{boussin99, lerot03, fejard03}. We used a broadening coefficient of 0.0613~\wn/atm (296/T)$^{0.5}$  for \dm~ colliding with a mixture of 86.3\%~\h~ and 13.6\%~helium, as measured by the Galileo Probe (von Zahn et al. 1998). Pressure-induced \h~ coefficients were obtained using laboratory measurements at 5 \um~ by \citet{bachet83} and the formalism developed by \citet{birnbaum76}.

The base of the model was set to 20 bars, 416K for the SEB Hot Spot to ensure that the base was well below the level where Jupiter's atmosphere becomes optically thick due to pressure-induced \h~ opacity. For the zone at 32\deg S, we investigated lower boundaries at 2, 4, 5, and 20~bars to simulate opaque \amhs~ clouds, \water~clouds, and no deep clouds in the troposphere. For the Hot Spot, we calculated a spectrum free of deep clouds for an emission angle of 18.7\deg~ and for an \water~ mole fraction of 47~ppm for $P>4.5$~bars as shown in Fig.~3, consistent with results from the Galileo Probe \citep{wong04}. The model was iterated to fit \dm~(0.18~ppm) and \ph~(0.45~ppm). Our fit to \dm~ in the Hot Spot is very close to the value (0.16$\pm$0.04~ppm) derived by \citet{lellouch01} using spectra at 8.6~\um~ from the Infrared Space Observatory. We then convolved the Jupiter spectrum to 0.02~\wn, doppler-shifted it by 25.2 km/sec, multiplied by the calculated transmittance above Mauna Kea, convolved it to 0.14~\wn to match the NIRSPEC data, and multiplied the spectrum by 0.332 to simulate the transmission of upper cloud layers on Jupiter to match the observed continuum. As shown in Fig.~4, the synthetic spectrum fits Region B, the SEB Hot Spot spectrum without any deep cloud opacity ($P>2$~bars). All 4 \dm~ features, including the wing between 2143.2 and 2143.6~\wn are matched.

The zone model is more complicated. A radiative transfer model was used to calculate spectra for an emission angle of 33.4\deg. Above each lower boundary, the gas composition was set to 0.18~ppm \dm. For the zone model with no deep clouds, the Galileo Probe \water~ profile was used. For models with opaque clouds at 2, 4, and 5~bars, a saturated profile of \water~ was used (see Fig.~3). The mole fraction of \ph~ was iterated to a value of 0.7~ppm to match the absorption feature at 2143 \wn for lower boundaries at 4 and 5~bars. For the zone model without deep clouds, the \ph~ abundance in the STZ was iterated to a value of 0.45~ppm. Thus, \ph~ can also be used to discriminate between deep cloud models, as we describe below. A larger \ph~ mole fraction (0.7~ppm) above an opaque cloud layer at 4 bars yields the same \ph~ absorption as a smaller mole fraction (0.45~ppm) above the (deeper) level where the atmosphere becomes opaque due to \h-\h~ opacity. The synthetic spectrum was split into two parts. The reflected solar component was calculated using the transmittances above a reflecting layer at 300~mbar and for an upper cloud reflectance of 9\%. It was convolved to 0.02~\wn resolution, doppler-shifted, and multiplied by the ATMOS solar spectrum. The thermal component was convolved, doppler-shifted, and multiplied by transmittances of 0.393, 0.0269, 0.016, and 0.0038 for models with a base of 2, 4, 5, and 20~bars, respectively, and added to the reflected spectrum. This, in turn, was multiplied by the transmission above Mauna Kea and finally convolved to 0.14~\wn resolution.

Due to the numerous parameters in this model, we illustrate the parameter sensitivity of the model for the reflected solar and thermal components of the zone model separately in Fig.~5. We treat the reflective upper cloud as spectrally grey. Ice components such as \ammonia, \amhs, and \water~ have been seen at other wavelengths (e.g., \citet{brooke98}, \citet{simon00}, \citet{baines02}, \citet{wong04b}, \citet{sromovsky10}). However, our spectral windows do not include significant ice absorption features. The imaginary indices of refraction of \ammonia~\citep{martonchik84} and \amhs~ \citep{howett07b} are on the order of $10^{-3}$ at the wavelengths studied here, much lower than values closer to unity at the $\nu_3$ vibration transitions for these ices (at 3.0 and 3.4 $\mu$m, respectively). Thus, the spectral features in the reflected solar component are due to \dm~ absorption lines in the upper troposphere of Jupiter or Fraunhofer lines in the Sun.

In the top panel of Fig.~5 we show 4 Fraunhofer lines, marked S. The strength of the strongest Jovian feature at 2141.8 \wn when compared with the spectrum of the Sun acquired by the ATMOS investigation yields the relative fractions of reflected sunlight (0.43) and thermal emission (0.57) for this portion of Jupiter's South Tropical Zone. The fraction of reflected sunlight combined with the calibrated continuum level  allows us to derive an upper cloud reflectance of 9\%. There are 5 \dm~ absorption features in this spectral range. Ammonia clouds and hazes limit the penetration of reflected sunlight to the upper troposphere. We investigated the pressure of the reflecting layer by calculating spectra at 100, 300, and 600 mbars. We adopted a value of 300 mbars for the STZ. Note that in order to match the narrow absorption cores of \dm~ in a model that consists of the sum of two components, there are a family of solutions for the pressure level of the solar reflecting layer and for the thermal cloud top pressure. Increasing the pressure of the reflecting layer from 300 to 600 mbars may be compensated by, for example, decreasing the (deep) cloud top pressure from 4 bars to 2 bars in the thermal component. However, we can exclude this possibility by studying the wing of the \dm~ feature between 2143.2 and 2143.6 \wnp. The top panel shows that the reflected solar spectrum is flat over this range. In contrast, each thermal model has a different \dm~ wing line slope. We now return to Fig.~4 which compares the sum of the reflected and thermal components of each model to the observed STZ spectrum. We can exclude models with an opaque \amhs~cloud at 2 bars (blue curve) as well as the model with no deep cloud (red curve). Note that the calculated radiances for both of these models lie well outside the error bars of the zone spectrum. The model with a deep cloud at 5 bars fits portions of the spectrum but the model with an opaque cloud at  4~bars (green curve) provides a better overall fit to the spectrum.

The NIRSPEC spectral bandpass for Order 16 covers more than just this particular \dm~ absorption feature. In Fig.~6 we compare the STZ spectrum with the same zone models for a spectral region adjacent to the one displayed in Figs. 4 and 5. Here there are three telluric \water~ lines. To the left of each telluric line is a Jovian water line red-shifted by 0.18 \wn due to the relative velocity of Jupiter with respect to the Earth. The wings of these water lines permit us to discriminate between models. The best fit requires an opaque cloud between 4 and 5~bars.

Next, we compare our retrieved \ph~ abundances in the 4 to 8-bar level of the STZ with measurements of \ph~ at 1 bar from the Cassini flyby of Jupiter in January 2001. \citet{irwin04} retrieved \ph~ mole fractions on Jupiter ranging from 0.9 to 1.5~ppm using zonal averages of CIRS spectra at 9 \um~ between 60\deg S and 60\deg N. The retrieved \ph~ mole fraction at 32\deg S was 1.0$\pm$0.2~ppm. Phosphine falls off with height in the upper troposphere due to ultraviolet photolysis, but its mole fraction is not expected to change between 1 and 8 bars. There remain discrepancies between \ph~ retrievals at 5 and 9 \um~ \citep{fletcher09}. Nevertheless, better agreement for the abundance of \ph~ at 32\deg S is achieved for a model with opaque clouds near 4 to 5 bars (0.7~ppm) than for a zone model with no deep cloud (0.45~ppm).

Thus, using three independent arguments that are based on: a) the slope of the \dm~ line wings, b) the slope of the \water~ line wings, and c) the derived mole fractions of \ph, we conclude that there must be significant cloud opacity between 4 and 5~bars in the STZ at 32\deg S. Based on the temperatures at these pressure levels (257K to 275K), thermochemical models predict that this is a water cloud.

In Fig.~7 we compare the spectrum of the SEB Hot Spot marked B at 2016~\wn (4.96~\um) with 3 different models calculated using the vertical profiles of \water~ shown in Fig.~3. The parameters are the same as those used to fit the spectrum at 2144~\wn except for \water~ and \ammonia. We initially used a vertical profile of \ammonia~ derived from absorption of the radio signal from the Galileo Probe (\citet{folkner98} as modified by \citet{hanley09}). We found that a scaling factor of 2 was required to fit the \ammonia~absorption features shown in Fig.~7. This corresponds to \ammonia~ mole fractions ranging from 240~ppm at 2~bars to 600~ppm at 4.5~bars. This is significantly larger than values derived from ground-based microwave observations (see \citet{sault04}). However, we have not used the much stronger absorption features in Order 14 (5.3 \um) to constrain the \ammonia~vertical profile. Thus, the \ammonia~ mole fractions reported here should be regarded as preliminary. The three different vertical profiles of \water~ in Fig. 3, and used in our calculations, were  based on the mass spectrometer data on the Galileo Probe.  \citet{niemann98} reported an upper limit to \water~ of 0.8~ppm at 2.7~bars. \citet{wong04} reported a mole fraction of \water~ of 40$\pm$13~ppm at 11.0-11.7~bars on Jupiter. The only adjustable parameter in our model was the pressure at which the \water~ mole fraction increased from 0.8 to 47~ppm, as shown in Fig.~3. The best fit is for a pressure of 4.5~bars. Assuming that the SEB Hot Spot is similar to the one that Galileo entered, this data point provides a useful measurement of the depth at which dynamical processes have dried out Hot Spots on Jupiter. This depth compares well with the depth determined by \citet{sault04} for the \ammonia~ abundance in hot spots from microwave observations.

We next modeled the zone marked A at 2016~\wnp. We used the same procedure as was used to model the zone spectrum at 2144~\wnp. We explored the same set of models with cloud tops at 2, 4, and 5 bars, as well as the model with no deep cloud. This portion of Jupiter's spectrum is not sensitive to the deep cloud structure. Using the results from fitting \dm, we adopted a model with an opaque cloud top at 4.0~bars and a saturated \water~profile above. The \ammonia~mole fraction was iterated until a rough fit was achieved using a value of 250~ppm between 0.75 and 4.0~bars and a saturated value for $P<0.75$~bars. Without any obvious Fraunhofer lines in this spectral region to constrain the reflected solar component, we assumed the same upper cloud reflectance of 9\% as at 2144~\wn and we obtained an upper cloud transmittance of 0.0542. The transmittance is therefore less certain than the value derived at 2144~\wnp. The principal conclusion from this spectral region is that the observed \water~ line profiles in the STZ are slightly broader than the telluric \water~features, but narrower than \water~features in the SEB Hot Spot. A saturated \water~profile for $P<4.0$~bars provides a satisfactory fit to the observed spectrum.

\begin{figure}[!ht]
\begin{center}
\begin{tabular}{ll}
\hspace{-0.12in}
	 \includegraphics[width=2.95in]{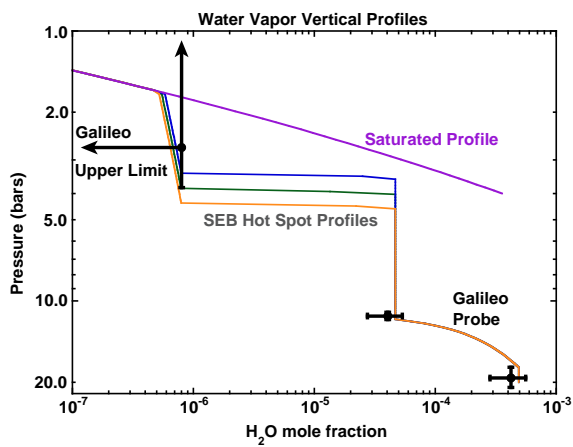}
\end{tabular}	\vspace{-0.1in}
   \caption {\footnotesize {Vertical profiles of \water~ used to calculate synthetic spectra in Figures 4 and 5.}}\vspace{-0.2in}
   \label{fig3}
\end{center}  
\end{figure}

\clearpage

\begin{figure*}[!ht]
\begin{center}
\begin{tabular}{ll}
\hspace{-0.15in}
	 \includegraphics[width=\textwidth]{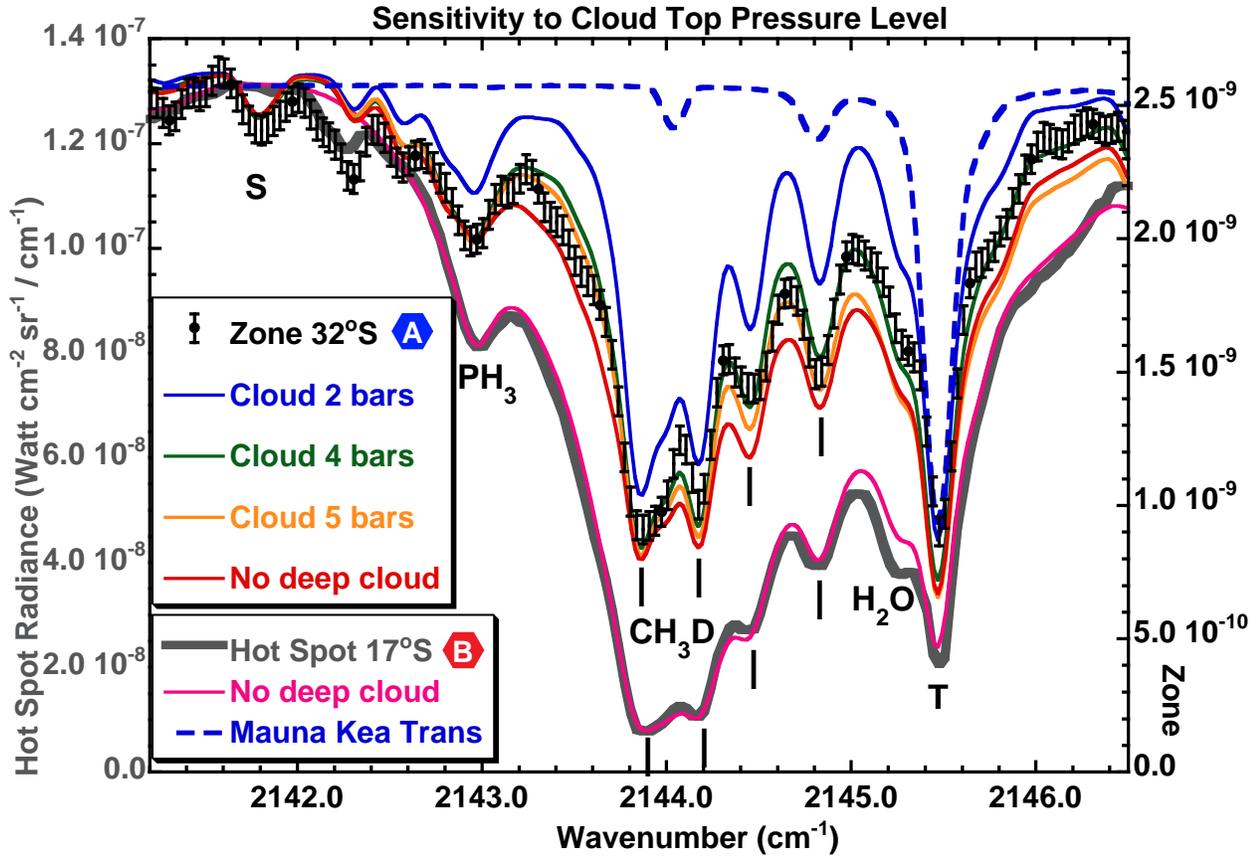}
\end{tabular} \vspace{-0.1in}
\caption  {\footnotesize {The pressure level of deep clouds on Jupiter as constrained by \dm~ line profiles observed in Order 16. An SEB Hot Spot (gray curve) at 17\deg S (labeled B in Fig.~1) is fitted without any clouds at pressures greater than 2 bars (pink curve). Hot Spot error bars would be smaller than the thickness of the gray curve. The spectrum of the STZ at 32\deg S (points with error bars, and labeled A in Fig.~1) was modeled using opaque clouds at 2, 4, 5 bars and a model without any deep clouds. We may exclude an opaque \amhs~ cloud at 2 bars due to the poor fit of the blue curve to the observed spectrum. The best fit requires an opaque cloud between 4 and 5~bars. Thermochemical models predict that cloud opacity at this pressure level is due to a water cloud. Note the factor of 50 difference in radiance scales between the Hot Spot and the zone spectra. The calculated transmittance above Mauna Kea for 2 mm precip \water~ is shown as a dashed blue line. T denotes a telluric \water~ line and S denotes a solar (Fraunhofer) line.}}
\vspace{-0.2in}
   \label{fig4}
\end{center}  
\end{figure*}  

\clearpage

\begin{figure*}[!ht]
\begin{center}
\begin{tabular}{ll}
	 \includegraphics[width=5.5in]{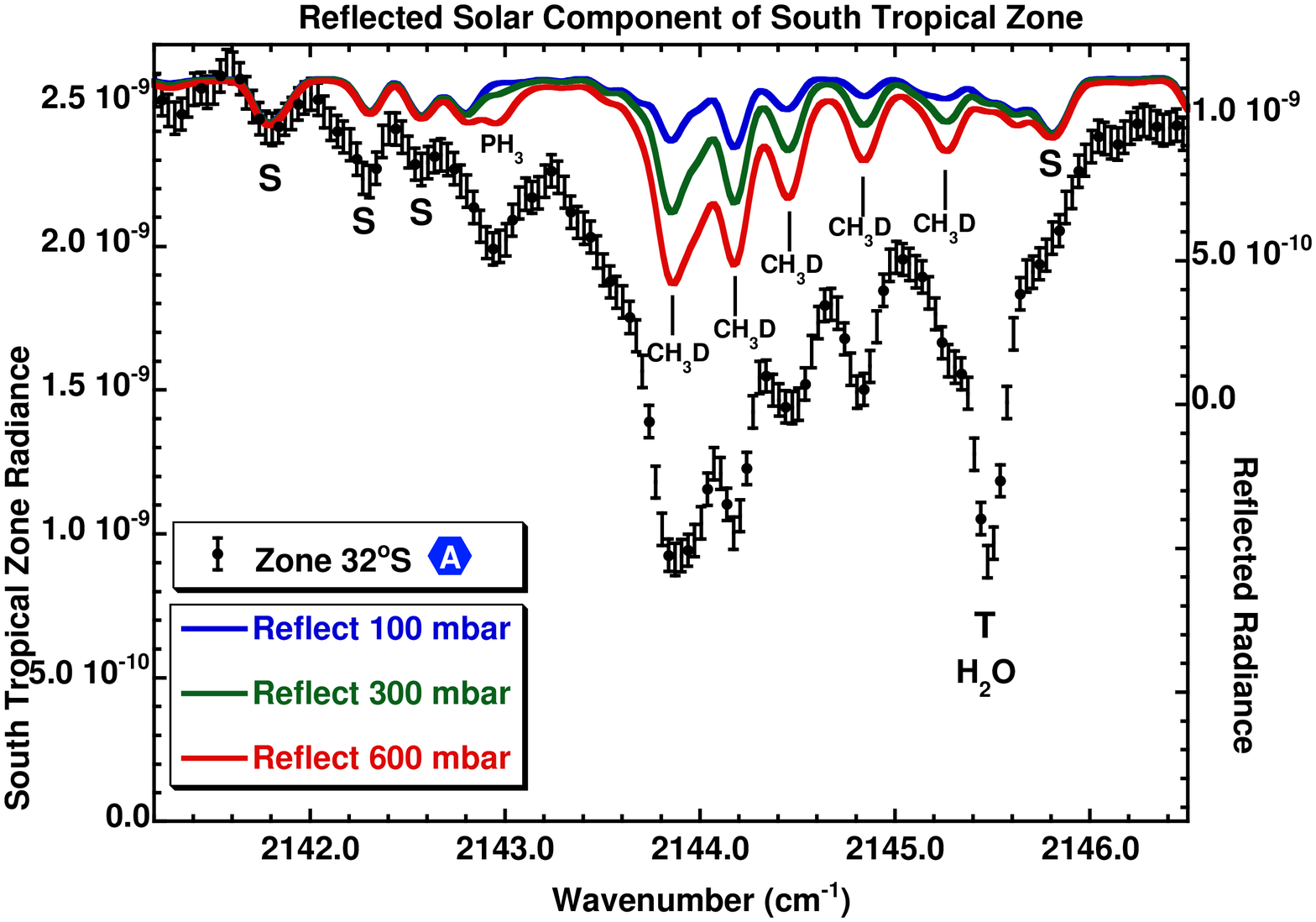}\\ 	  
	 \includegraphics[width=5.5in]{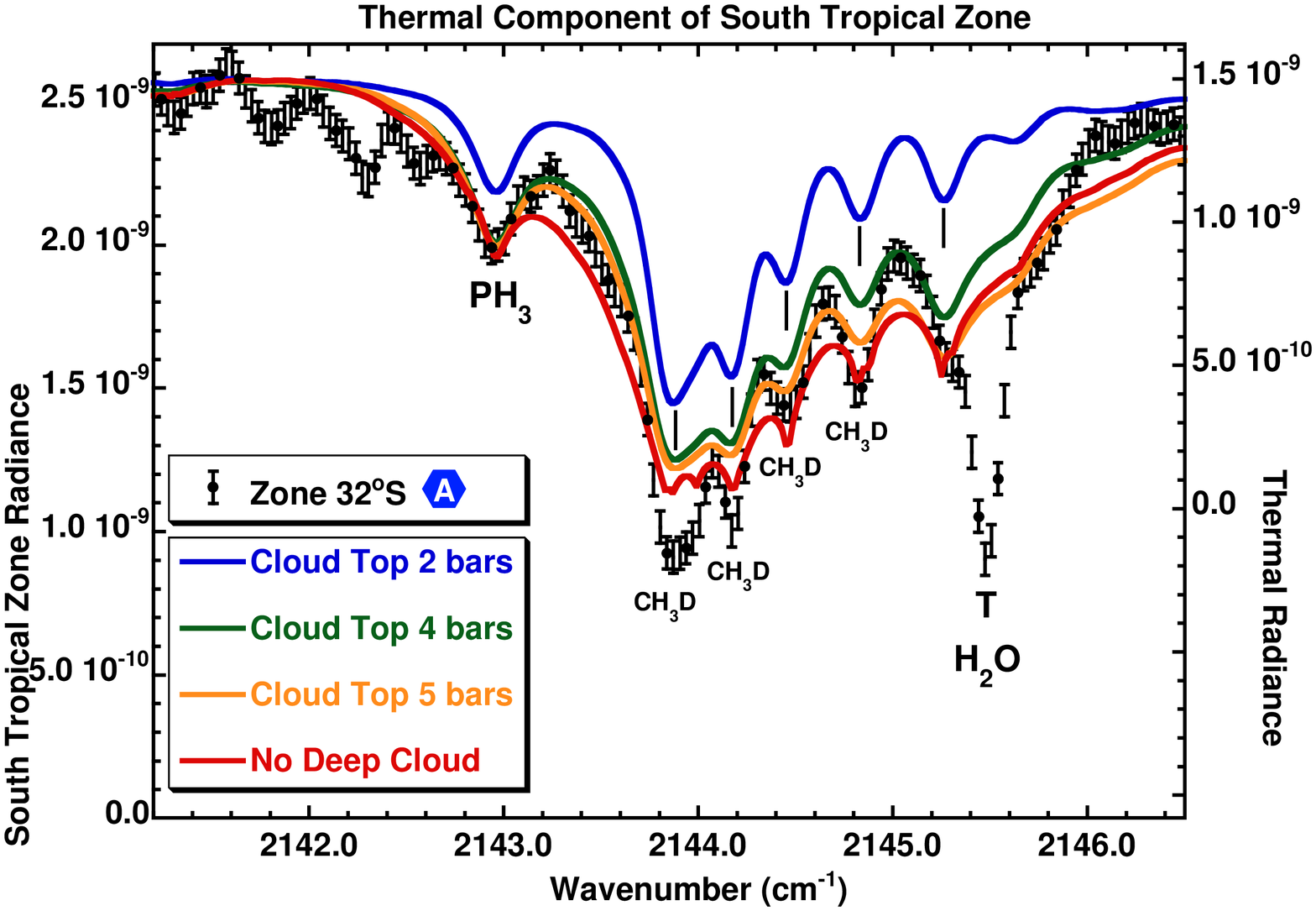}
\end{tabular} \vspace{-0.1in}
\caption  {\footnotesize {Top: The reflected solar component of a radiative transfer model of the STZ at 32\deg S. Four Fraunhofer lines are marked as S. A telluric \water~ line is marked as T. A reflecting layer is placed at 100, 300, and 600 mbars in Jupiter's upper troposphere. One \ph~ and five \dm~ lines are shown. The narrow cores of the \dm~ lines are sensitive to the solar component. Bottom: The thermal component is shown for models with cloud tops at 2, 4, and 5~bars, and for a model with no deep cloud. The wing of \dm~ between 2143.2 and 2143.6 \wn and the \ph~ line at 2143.0 \wn are due to the thermal component. Radiance scales are offset to match the continuum of the observed spectrum.}}
\vspace{-0.25in}
   \label{fig5}
\end{center}  
\end{figure*}  

\clearpage

\begin{figure*}[!ht]
\begin{center}
\begin{tabular}{ll}
\hspace{-0.15in}
	 \includegraphics[width=\textwidth]{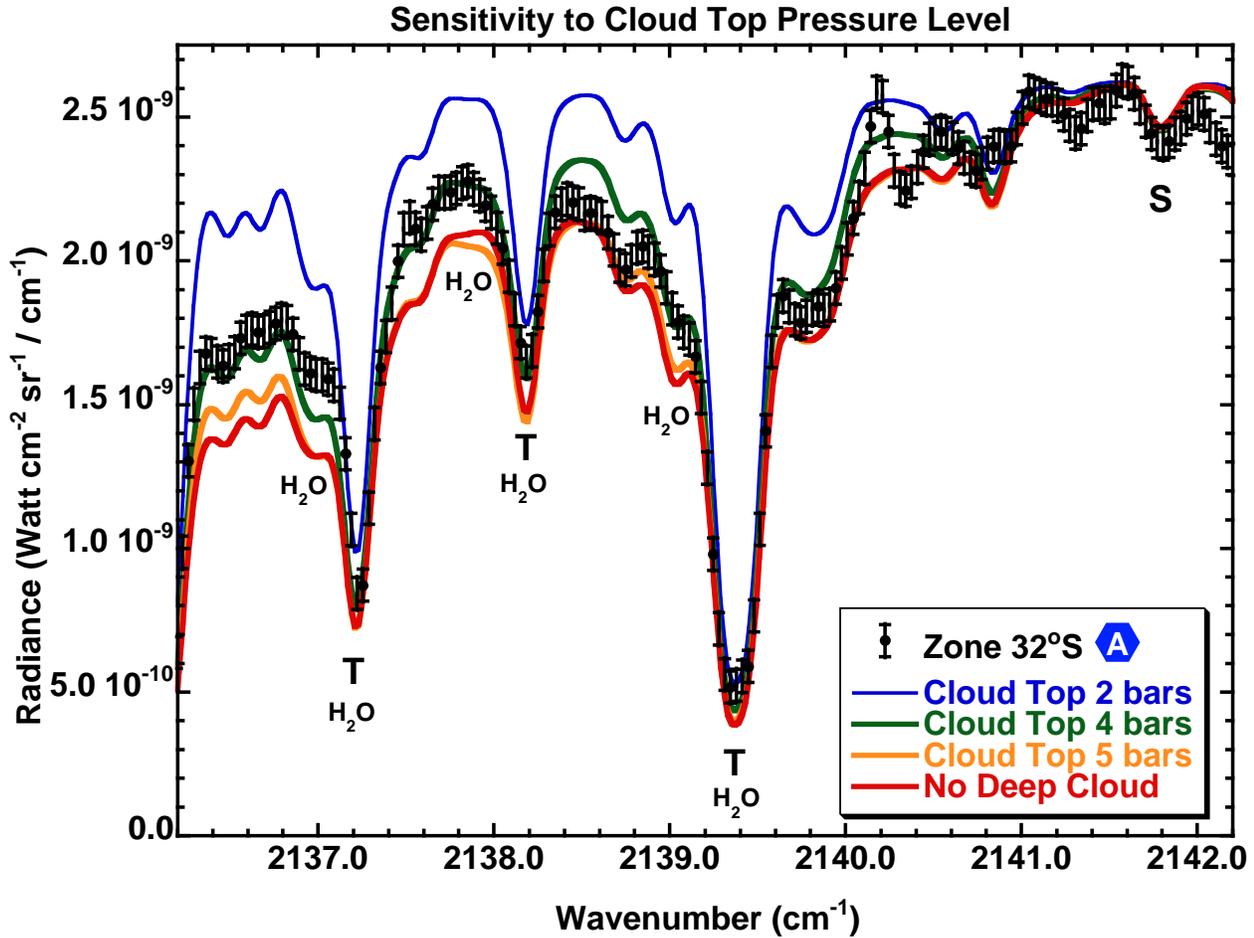}
\end{tabular} \vspace{-0.1in}
\caption  {\footnotesize {The spectrum of the STZ at 32\deg S and radiative transfer models with opaque clouds at 2, 4, and 5 bars, and a model without any deep clouds are shown for a spectral region adjacent to the \dm~ features.  T denotes telluric \water~ lines and S denotes a solar (Fraunhofer) line. Note the 3 Jovian \water~ lines red-shifted by 0.18 \wn from their telluric counterparts. The wings of these water lines permit us to discriminate between models. The best fit requires an opaque cloud between 4 and 5~bars.}}
\vspace{-0.2in}
   \label{fig6}
\end{center}  
\end{figure*}

\clearpage

\begin{figure*}[!ht]
\begin{center}
\begin{tabular}{ll}
\hspace{-0.15in}
	 \includegraphics[width=\textwidth]{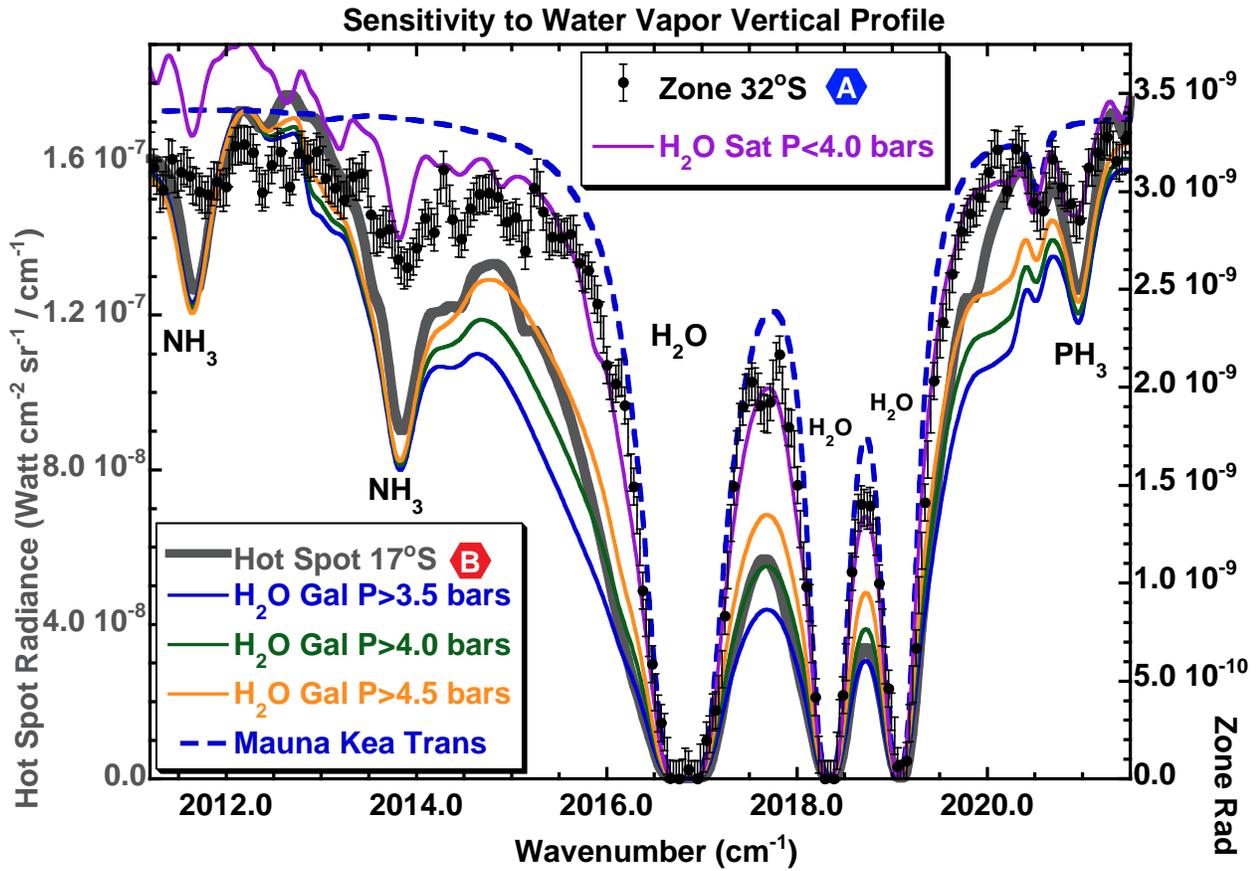}
\end{tabular}	\vspace{-0.1in}
   \caption {\footnotesize {Vertical profiles of water vapor mixing ratio as constrained by \water~ line profiles observed in Order 15. An SEB Hot Spot (Region B, gray curve) at 17\deg S is fitted without any deep clouds. Hot Spot error bars would be smaller than the thickness of the gray curve. Three vertical profiles of \water~ were calculated for the Hot Spot. The best fit used the Galileo Probe value for \water~ for pressures greater than 4.5 bars (see also Fig.~3). The spectrum of the STZ at 32\deg S (Region A, points with error bars) was fitted using an opaque cloud at 4 bars and a saturated \water~ profile above it (purple curve). The \water~ lines in the zone are slightly broader than the telluric features, but much narrower than in the Hot Spot.}}\vspace{-0.2in}
   \label{fig7}
\end{center}  
\end{figure*}

\clearpage

\begin{figure*}[!h]
\centering{
\begin{tabular}{ll}
\hspace{-0.15in}
	 \includegraphics[height=4.0in]{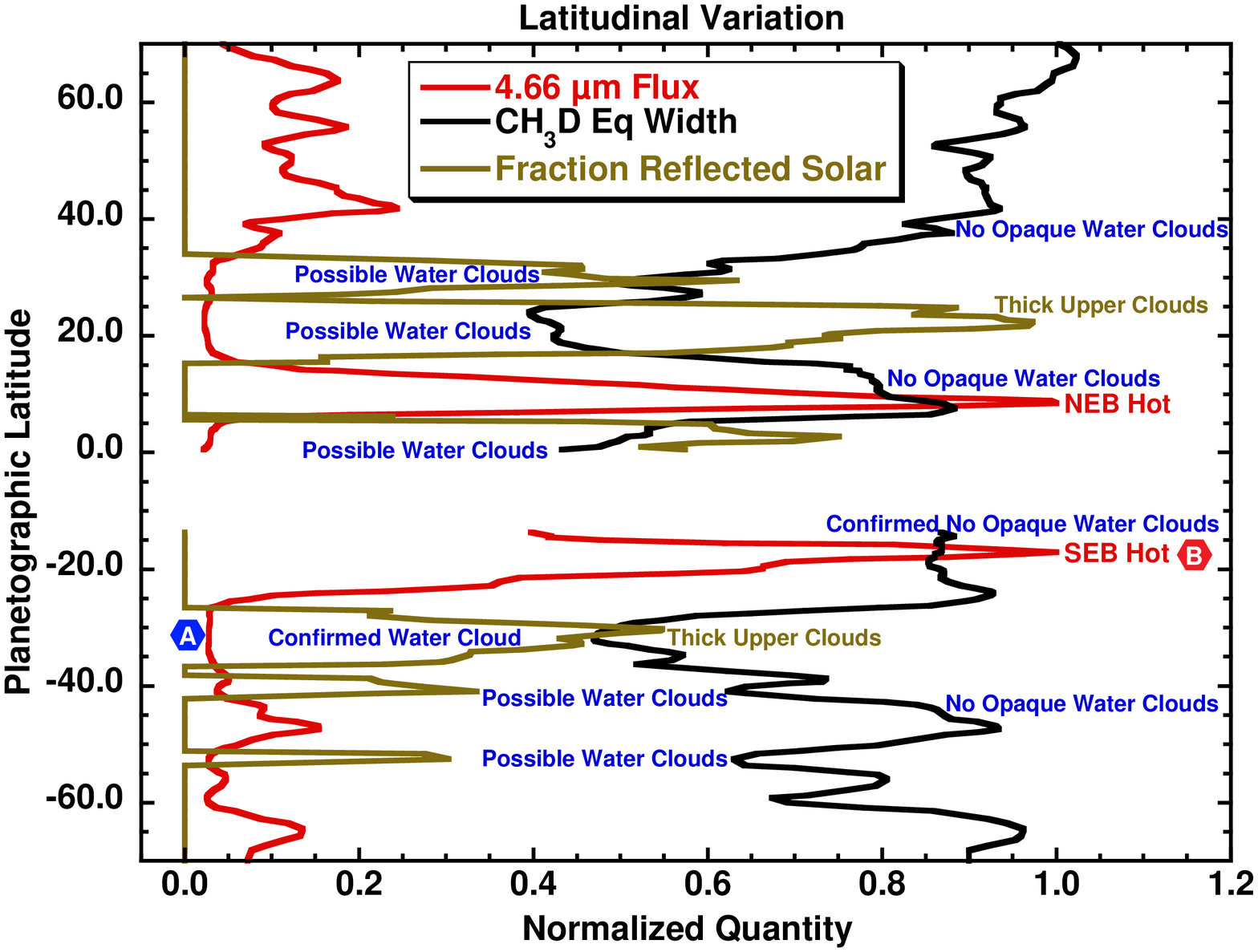} &
	\hspace{-0.2in}\includegraphics*[bb=0in -1.13in 1.in 15in, scale=0.352]{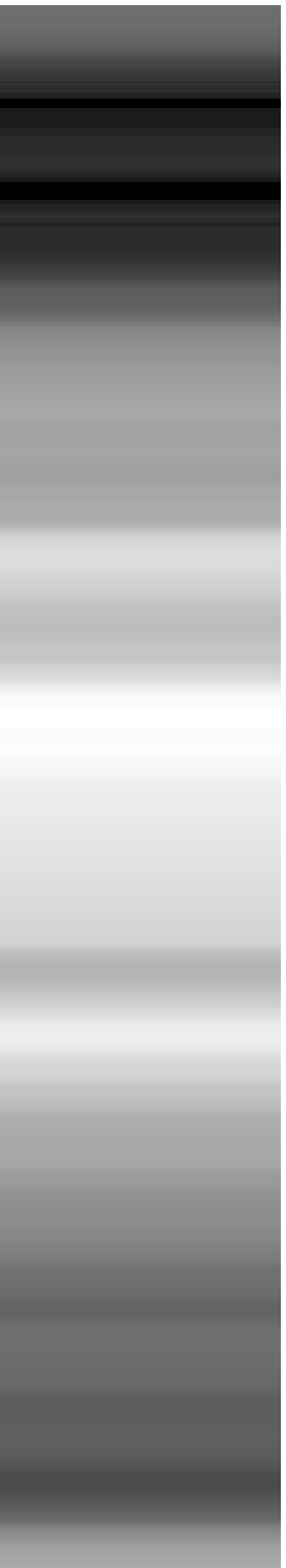}
\end{tabular}	\vspace{-0.1in}
   \caption {\footnotesize {The flux at 4.66 \um~ (red curve), the equivalent width of \dm~ (black curve), and the the ratio of the equivalent width of the Fraunhofer line at 2141.8 \wn on Jupiter to that in the Sun is shown for each latitude along the central meridian on Jupiter. Hot Spots are present at 8.5\deg N and 17\deg S. Latitudes exhibiting minima in \dm~ equivalent widths are candidates for water clouds. Radiative transfer models were used to match a zone spectrum at 32\deg S (marked "A") and a Hot Spot spectrum at 17\deg S (marked "B"). The gold curve indicates the fraction due to reflected sunlight. Large values imply thick upper clouds that attenuate thermal radiation from below and reflect sunlight above. The bar on the right denotes haze reflectivity variation with latitude obtained from 2.12~\um~ SCAM images, roughly similar to belt-zone structure seen at visible wavelengths.}}\vspace{-0.2in}
   \label{fig8}
}
\end{figure*}

\clearpage

\section{DISCUSSION}

This paper describes a technique to determine water cloud heights using spectrally-resolved line shapes of \dm. We have demonstrated with a set of spectra from characteristic spatial regions that variation in the water cloud height manifests as a variation in the equivalent width of \dm. This technique can be applied to all regions covered by the spectrometer slit. We show the \dm~ equivalent width as a function of latitude (black curve) in Fig. 8, highlighting the latitude regions which may have water clouds. The observations are extremely valuable both for understanding Jupiter's atmospheric circulation, as well as its bulk water abundance (a constraint on planetary formation scenarios).

However, the equivalent width of \dm~ alone is not sufficient. In order to map \water~ clouds on Jupiter, we need to measure the latitudinal profile of both \dm~ line shapes as well as the strength of the Fraunhofer line at 2141.8~\wnp. In Fig.~8 we plot three quantities as a function of latitude on Jupiter. First is the radiance integrated over the 4.6-\um~ bandpass and normalized to 1.0 as a function of spatial pixel and converted to planetographic latitude. This is shown in red. The next quantity is the equivalent width of our set of \dm~ absorption lines integrated between 2143.21 and 2145.04~\wnp (denoted by the horizontal bar in Fig.~2). This is shown in black. The last quantity plotted in Fig.~8 is the fraction of reflected sunlight as a function of latitude. This was obtained by measuring the equivalent width of the Fraunhofer line at 2141.8~\wn as a function of latitude and dividing it by the value in the Sun as measured by ATMOS. At latitudes such as 23\deg N where this value is close to 1.0, there is insufficient thermal flux to constrain the deep cloud structure. However, by using adjacent latitudes where \dm~ equivalent widths are also small, and where the fraction of reflected solar flux $<$ 0.6, we can infer the presence of water clouds. Thus, we have indicated "possible water clouds" for the entire region where values of \dm~ equivalent width are small. The vertical bar on the right denotes the brightness of Jupiter in reflected sunlight at 2.12~\um~obtained from SCAM images navigated onto a planetographic latitude grid. This wavelength sounds hazes in the upper troposphere and illustrates Jupiter's belt-zone structure.

The equivalent width of \dm~ is sensitive to two parameters: the fraction of reflected sunlight and the pressure level of the deep cloud. Consider two regions on Jupiter. They both lack an opaque water cloud, but one has thin upper clouds so that the 5-\um~ spectrum is 100\% thermal. The other has a thick, reflective upper cloud that attenuates the thermal component yielding a spectrum that is 50\% thermal and 50\% reflected solar. In the latter case, the contribution of the reflected solar continuum will significantly reduce the equivalent width of the \dm~ absorption features that are present in the thermal component. A low value of \dm~ equivalent width might be interpreted as due to a deep water cloud, whereas in this case the small equivalent width was due solely to the presence of the solar component from an upper reflective layer. Thus, we have to be cautious in using the equivalent width of \dm~ alone to infer deep cloud structure. Nevertheless, the data plotted in Figure 8 show interesting structure at latitudes where the fraction of reflected solar radiation is essentially zero. High latitudes, southward of 55\deg S and northward of 35\deg N show interesting variations in \dm~ equivalent widths that cannot be attributed to reflected sunlight and thus are related to deep cloud structure.

With the caveat that radiative transfer modeling is required to verify the presence of water clouds in regions with small \dm~ equivalent widths and significant fractions of reflected sunlight, we can make the following conclusions: Where thick water clouds are present at pressures in the 3-5 bar range, the \dm~ equivalent width is small, and the black curve (Fig. 8) tends toward lower values. Regions with thick, high-altitude ($P \leq 4$~bars) water clouds can be associated with widespread upwelling, since downwelling would provide the opposite effect of clearing cloud opacity via sublimation. We find the smallest \dm~ equivalent widths in the three low-latitude zones: the equatorial zone and the north and south tropical zones. This implies a large vertical extent to the upwelling responsible for forming the thick white clouds observed in zones at optical wavelengths.

Thick water clouds in zones specifically contradict models of inverted two-layer circulation within the tropospheric cloud decks. \citet{ingersoll00} and \citet{showman05} suggested inverted two-layer circulation schemes to explain widespread ammonia gas depletion even in zones. In this type of circulation model, regions of upwelling (thick clouds) in the visible upper levels (\ammonia + \amhs~ cloud decks) corresponds to downwelling at the deeper, hidden water cloud level. Our results suggest that mass flux into the upper cloud layers of zones is not dominated by horizontal transport, as in the  \citeauthor{showman05} scenario, but is driven by vertical transport from below. Jupiter's circulation in zones therefore maintains the same sign of upwelling/downwelling across the full 0.5-5~bar weather layer. The same sign of upwelling/downwelling over this large an extent was also derived by \citet{depater10} from 5-micron bright rings around vortices. These authors suggested that vortices must extend vertically from at least the 4-7 bar level up to the tropopause.

Large \dm~ equivalent width requires low water cloud opacity. Subsidence of dry upper-level air provides the simplest mechanism. Figure 8 shows that at low latitudes, the clearest deep atmosphere (high values of the black curve) occur where the 5-$\mu$m flux is highest (red curve). The very highest flux levels correspond to an atmosphere largely devoid of clouds of all three types, \water, \ammonia, and \amhs, again suggesting downwelling circulation that spans the full 0.5-5 bar range at least. However, the width (in latitude) of the 5-$\mu$m flux peak is very narrow, compared to the width of the NEB/SEB regions with low water cloud opacity. Thus, over a broad latitude range with little or no water cloud opacity, the atmosphere is characterized by both cloudy and cloud-free conditions at the upper levels. Perhaps a two-layer circulation model with a vertical flow reversal layer between the water and \ammonia / \amhs~ clouds can be relevant on these regional scales, if not for the entire planet.

The rapid ``fading" or whitening of the South Equatorial Belt between 2008 and 2010 was interpreted by \citet{fletcher11a} as due to enhanced upwelling of ammonia-rich air followed by condensation. Using \dm~ line profiles and absorption lines of \ammonia~ and \water~ in the 5-\um~window, we now have the capability of measuring cloud structure and volatile abundances in the NEB and SEB at the 4 to 8~bar level.  Any future changes in the appearance of the belts can now be investigated over the full 0.5 to 8-bar range of Jupiter's troposphere using spectroscopy at 4.6 and 8.6~\um. By studying changes in each cloud layer separately we will have a much better understanding of the dynamics below Jupiter's visible clouds.

The high latitudes also show high \dm~ equivalent widths. Here, more detailed modeling will be needed to disentangle geometric effects caused by viewing geometry, as well as changes in atmospheric scale height in a rapidly-rotating nonspherical planet. But the data do suggest a different paradigm in deep cloud structure at high latitudes (polewards of 40\deg ). We note, though, that 5-micron images at high spatial resolution show a lot of structure at these latitudes \citep{depater11}, and microwave images show an overall low \ammonia~abundance \citep{depater86}. A combination of 5-\um~ imaging, microwave imaging, and 5-\um~spectroscopy of Jupiter's polar regions would be extremely useful to investigate these interesting cloud features.

We now return to the interpretation of 5-\um~ spectra of Jupiter acquired by the KAO, Voyager/IRIS, and Galileo/NIMS. It now seems clear that the airborne observations of Jupiter were flux-weighted by Hot Spots. The abundances of \water~ and other molecules derived by \citet{bjoraker86b, bjoraker86a} pertain to Hot Spots, but not to Jupiter as a whole. The lower boundary for Hot Spots is in fact due to pressure-induced \h~ opacity, rather than an opaque water cloud, as proposed by \citet{carlson92}. However, the model proposed by \citeauthor{carlson92} does appear to apply to Jupiter's zones, at least in the regions marked in Fig.~8 as candidates for water clouds. Similarly, models of Galileo/NIMS spectra of the Equatorial Zone by \citet{drossart98} are probably accurate while models of Hot Spots that included water clouds (e.g. \citet{nixon01}) will need to be revised.

This technique provides a constraint on Jupiter's deep water abundance and therefore its O/H ratio. The base of the water cloud is sensitive to the abundance of water because higher abundances lead to condensation at deeper levels. The data provide the level of the cloud top, not the cloud base. Since the top is at higher altitude than the base, cloud top constraints provide lower limits to the pressure of the cloud base, or, lower limits to the deep abundance of water. A spectrum requiring a water cloud at $P\ge5$~bar would establish a supersolar enrichment of water in Jupiter, better constraining planetary formation models (Wong et al. 2008). 

Figure~4 shows the effect of water cloud pressure level on model fits to a cloudy zone spectrum at 32\deg S. The spectrum is best fit by a water cloud with a top between 4-5 bars. The cloud base is therefore found at $P > 4-5$~bar. Following Fig.~1 in \citet{wong08}, a cloud base at 4-5 bar corresponds to O/H ratios 0.33-1.1$\times$ solar (corrected to the new solar O/H ratio of Asplund et al. 2009). Our observations thus provide a lower limit to Jupiter's water abundance of 0.33-1.1$\times$ solar. This result is consistent with the Galileo Probe lower limit of 0.48$\pm$0.16$\times$ solar (Wong et al. 2004), and therefore does not provide any new constraint on the water abundance. In future work, we will search for spectra that require an even deeper water cloud. 

Knowledge of the deep cloud structure permits us to retrieve abundances of \water, \ammonia, and other 5-\um~ absorbers more accurately. This will enable studies of dynamics below Jupiter's visible cloud layers. Additional observations will permit us to observe discrete cloud features such as the Great Red Spot and Oval BA.



\acknowledgments

The data presented were obtained at the W. M. Keck Observatory, which is operated as a scientific partnership among the California Institute of Technology, the University of California, and the National Aeronautics and Space Administration.  The Observatory was made possible by the generous financial support of the W. M. Keck Foundation. The authors extend special thanks to those of Hawaiian ancestry on whose sacred mountain we are privileged to be guests. Without their generous hospitality, none of the observations presented would have been possible. We also would like to thank Linda Brown for steering us to the latest broadening coefficients for \dm. This research was supported by  the NASA Planetary Astronomy (PAST) Program  grant number NNX11AJ47G,  NNX14AJ43G, NNX15AJ41G, and NASA Outer Planets Research  Program grant number NNX11AM55G.



{\it Facilities:} \facility{Keck Observatory}, \facility{Voyager (IRIS)}.
\bibliography{Jupiter_Saturn.bib}
\bibliographystyle{apj}


\clearpage



\end{document}